\documentclass[comsoc,journal]{IEEEtran} 

\usepackage{cite,graphicx,amsmath,amssymb}
\usepackage{comment}
\usepackage{graphicx}
\usepackage{amsfonts,amssymb}
\usepackage{xcolor}
\usepackage{multirow}
\usepackage{booktabs}
\usepackage{subcaption}
\usepackage{url}
\usepackage{enumerate}
\usepackage{enumitem}
\usepackage[
top    = 1.70cm,
bottom = 0.82 in,
left   = 0.63 in,
right  = 0.63 in]{geometry}
\usepackage{xspace}
\usepackage{diagbox}

\newcommand\jiadong[1]{\textcolor{black}{#1}}

\title{Network Topology and Information Efficiency of Multi-Agent Systems: Study based on MARL}
\author{Xinren Zhang, Sixi Cheng, Zixin Zhong, Jiadong Yu
\vspace{-.7cm}
\thanks{X. Zhang, Z. Zhong, and J. Yu are with the Hong Kong University of Science and Technology (Guangzhou).}
\thanks{S. Cheng is with the Singapore University of Technology and Design.}

}
\vspace{-.1cm}
\begin{document}
\maketitle
\begin{abstract}
Multi-agent systems (MAS) solve complex problems through coordinated autonomous entities with individual decision-making capabilities. While Multi-Agent Reinforcement Learning (MARL) enables these agents to learn intelligent strategies, it faces challenges of non-stationarity and partial observability. Communications among agents offer a solution, but questions remain about its optimal structure and evaluation. This paper explores two underexamined aspects: communication topology and information efficiency. We demonstrate that directed and sequential topologies improve performance while reducing communication overhead across both homogeneous and heterogeneous tasks. Additionally, we introduce two metrics -- Information Entropy Efficiency Index (IEI) and Specialization Efficiency Index (SEI) -- to evaluate message compactness and role differentiation. Incorporating these metrics into training objectives improves success rates and convergence speed. Our findings highlight that designing adaptive communication topologies with information-efficient messaging is essential for effective coordination in complex MAS.
\end{abstract}
\vspace{-.3cm}
\section{Introduction}
\label{sec:intro}
Multi-agent systems (MAS) comprise collections of autonomous computational entities that perceive, reason, act, and interact to achieve individual or collective objectives in shared environments~\cite{wooldridge2009introduction}. These systems provide a powerful paradigm for modeling complex phenomena across domains, from social dynamics to distributed computing and robotic coordination. Multi-Agent Reinforcement Learning (MARL) has emerged as a cornerstone of modern artificial intelligence, offering a principle framework for distributed decision making in environments where multiple autonomous agents interact~\cite{gronauer2022multi}. These agents may cooperate to achieve shared goals, compete against each other in adversarial settings, or operate in mixed scenarios where collaboration and competition coexist. Such flexibility makes MARL a powerful tool for a wide range of applications, from drone swarms and robotic teams to intelligent transportation systems and smart grids. By enabling agents to learn from interaction and to adapt to uncertainty, MARL provides the foundation for resilient, and adaptive MAS. However, this promise comes with significant challenges, as the presence of multiple decision-makers inherently introduces non-stationarity of environment to each single agent (where the environment appears to change as other agents update their policies), coordination complexity, and communication constraints.

One of the fundamental difficulties in MARL is that each agent has only partial observations of the environment and needs to adapt in a dynamic setting where other agents are simultaneously updating their strategies. This non-stationarity severely complicates the learning process. The widely adopted Centralized Training with Decentralized Execution (CTDE) paradigm addresses some of these challenges by using global information during training while relying on local observations during execution. 
However, CTDE alone cannot fully resolve the limitations of partial observability, especially during the execution stage. It also faces challenges in scaling to large, heterogeneous MAS, which consist of agents with different capabilities, observation spaces, action spaces, and reward functions.

To overcome these obstacles, researchers have increasingly turned to structuring the communications in MARL\cite{kim2021communication}. By allowing agents to exchange information, communication extends the effective observation space of each agent, reduces uncertainty about the environment state and other agents' behaviors, and enables richer forms of cooperation. However, communication itself raises new design questions: \textit{with whom should an agent communicate, when should communication occur, and what information should be transmitted?} These questions point directly to two core aspects of communication design -- the network topology that governs how information flows among agents, and the information efficiency that determines the usefulness of the exchanged messages. This paper investigates these two aspects in depth with following two Research Questions (RQs).

\textbf{\jiadong{RQ1: How would network topology influence MARL performance?}}

In MARL, the communication network topology defines both the connectivity patterns between communicating agents and the order in which messages propagate. Early approaches often relied on full broadcasting, which maximizes observability but introduces redundancy and inefficiency. In this paper, we explore the impact of the network topology in MARL from two complementary angles: the direction of communication links and the order of information flow. The direction of links determines connectivity constraints and hierarchical relationships, while the sequence of message propagation allows later agents to adapt to preceding agents' actions, mitigating the effects of non-stationarity and improving coordination. Understanding how directed connectivity and sequential propagation interact based on both homogeneous and heterogeneous agent settings is key to advancing MARL toward more adaptive and efficient MAS.

\textbf{\jiadong{RQ2: How would the communication information influence the MARL performance?}}

While RQ1 emphasized the role of network topology in determining the communication connectivity patterns among agents and the sequential propagation order of information, an equally important question is what information is actually exchanged. The efficiency of communication depends not just on the quantity of messages but on their compactness. Early approaches such as hidden-state sharing enabled differentiable message passing, but they often introduced redundancy and bottlenecks as system scale increased. Attention-based methods addressed this challenge from two angles: targeted attention ensures that agents selectively interact with relevant peers, while attention frameworks further refine message importance. To systematically capture this efficiency, we employ the Information Entropy Efficiency Index  (IEI) and Specialization Efficiency Index (SEI), which measure the ratio of the entropy and the similarity of the transmitted messages to task performance, respectively. A lower IEI indicates that agents are conveying more task-relevant information with less redundancy. A lower SEI indicates that agents have distinct roles. By integrating IEI and SEI into the training process, MARL systems can be guided toward compact and diverse communication protocols that not only reduce overhead but also accelerate convergence and improve coordination effectiveness.

Thus, by linking network structural design and message quality, this paper emphasizes that robust MARL requires not only well-designed communication topology but also efficient use of communication channels, where agents learn what to share, when, and with whom to share it.

The paper is organized as follows: Section~\ref{sec:emergent_comm} introduces communications in MARL and the general framework. Section~\ref{sec:network} examines network topology in MARL through a case study demonstrating how connectivity, directionality, and sequential propagation enhance task efficiency. Section~\ref{sec:comm} explores agent communication mechanisms, presenting metrics for communication efficiency that improve training performance. Section~\ref{sec:challenge} discusses limitations and future research directions, while the final section concludes the paper.

\section{Communications in MARL}
\label{sec:emergent_comm}
This section elucidates the theoretical foundations and architectural paradigms of communication mechanisms in MARL systems, with particular emphasis on their structural implementation and functional dynamics.
\vspace{-.4cm}
\subsection{MAS and MARL}
MAS model complex interactive phenomena through distributed autonomous entities that perceive their environment, make decisions, and coordinate to achieve goals~\cite{wooldridge2009introduction}. Reinforcement Learning (RL) provides a mathematical framework in which an agent learns to make sequential decisions by interacting with its environment, receiving rewards as feedback, and optimizing its policy to maximize long-term returns. MARL extends the foundational principles of RL to MAS.
Depending on the system architecture and accessibility of information, MARL approaches are broadly categorized into CTDE and Fully Decentralized Learning~\cite{gronauer2022multi}. In the CTDE framework, agents are trained using centralized information - such as the global state and joint actions of all agents - which simplifies coordination and allows for more efficient credit assignment. However, during execution, each agent must act independently based only on its local observations. In fully decentralized learning, all centralized components are removed. Each agent learns its own policy based solely on locally available information and receives rewards. While this method enhances autonomy and suits large-scale systems or communication-constrained environments, it introduces severe learning challenges. Foremost among these challenges is the non-stationarity problem, wherein the adaptation of other agents' policies creates an unstable learning environment from the perspective of any individual agent. This environmental instability significantly impedes convergence during training and often results in suboptimal policy solutions.

\begin{figure}[t]
  \centering
  \includegraphics[width=.37\textwidth]{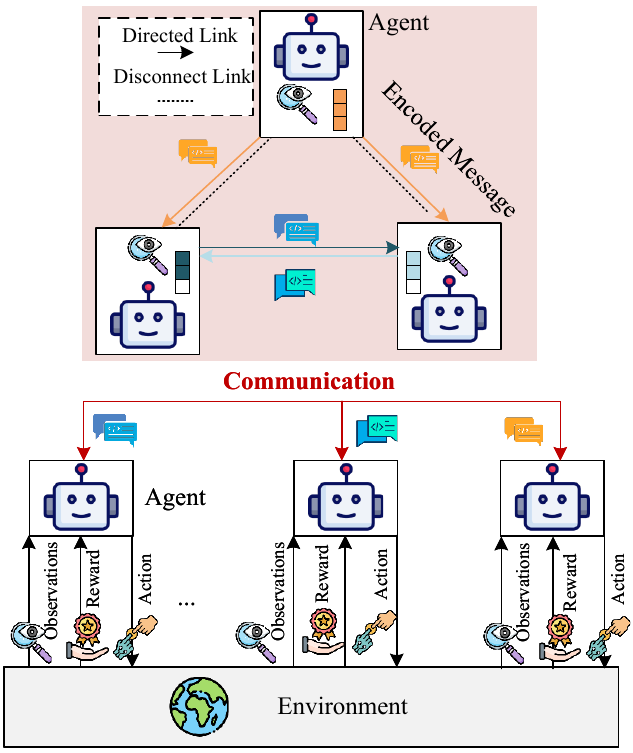}
  \caption{Illustration of the communications in MARL.}
  \label{fig1}
  \vspace{-.5cm}
\end{figure}
\vspace{-.35cm}
\subsection{Communications in MARL}

The limitations discussed above, especially partial observability and non-stationarity, have led to a growing interest in structuring communication within MARL systems. Communication allows agents to learn to exchange meaningful representations through communication channels, which are refined during training rather than predefined~\cite{kim2021communication}. This approach enables agents to dynamically develop communication protocols that enhance collective decision-making and task performance.

Fig.~\ref{fig1} illustrates how communications expands the effective observation space of each agent. Instead of relying solely on their local views, agents learn to broadcast or selectively send messages that convey relevant information of their own observations to others. This additional interaction improves situational awareness, reduces uncertainty, and fosters coordinated behaviors. This learned communication mechanism is particularly powerful in scenarios with heterogeneous agents or non-stationary objectives, where pre-defined communication rules may not generalize well. It offers a way to flexibly coordinate agents in real-time, even as the environment or agent group composition evolves. 
\vspace{-.4cm}
\subsection{General Framework}

\begin{figure*}[t]
  \centering
  \includegraphics[width=.87\textwidth]{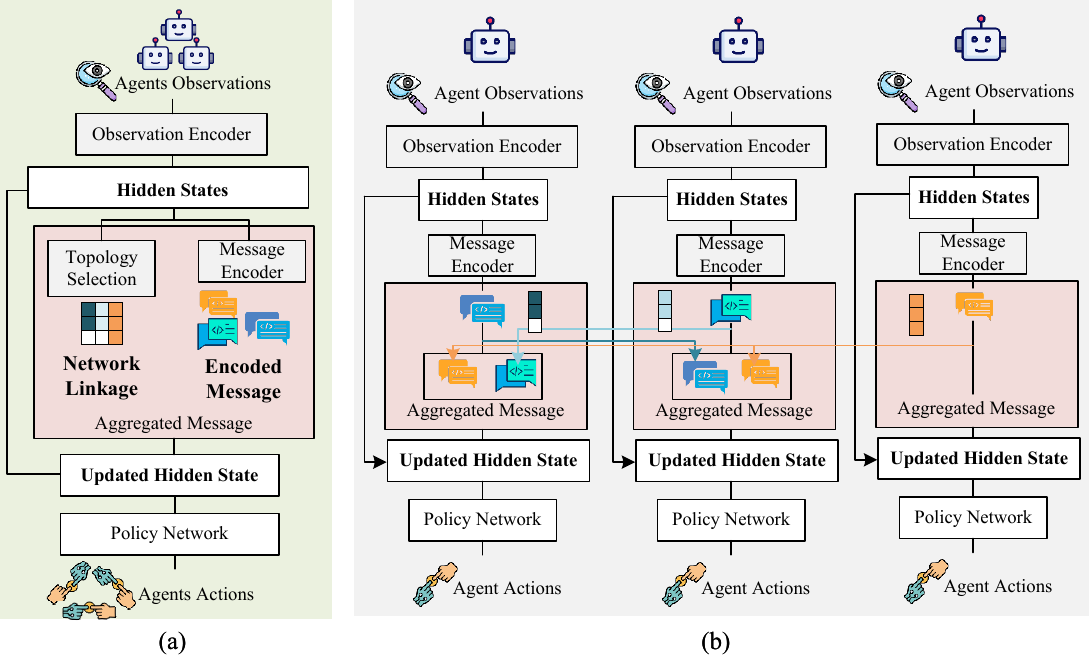}
  \caption{General framework of the CTDE paradigm of the MARL with communications. (a) Centralized training, and (b) Decentralized execution.}
  \label{fig2}
  \vspace{-.4cm}
\end{figure*}

To enable effective coordination in complex environments, MARL with communications can be naturally integrated with the CTDE paradigm. This framework enables agents to learn sophisticated communication strategies during training, while maintaining full autonomy and decentralization during execution.
As shown in Fig.~\ref{fig2}, the framework operates in two distinct stages:

\subsubsection{Centralized Training Stage}
In this stage, a central controller with access to global observations guides the joint learning process of all agents' decision-making policies and their communication protocols. Each agent first processes its local observations through an encoder to generate internal feature representations. These features are then encoded into messages, which serve as the distilled \textit{information} representation of each agent’s local state.

Another critical component of the training process is the communication \textit{topology}, a dynamic graph that determines communication connectivity. This topology may be predefined or learned as part of the training process. Once established, each agent aggregates messages from its connected neighbors in the communication graph. This aggregated information is then used to update the agent’s internal representation, refining its understanding of the environment and other agents’ intentions. Finally, using these enriched representations, the agents’ policies are refined with the help of a centralized critic that evaluates joint performance. The training process thus encourages agents to develop communication behaviors that improve overall cooperation.

\subsubsection{Decentralized Execution Stage}
During execution, each agent acts independently using its own local observations and the learned communication protocol. The previously trained communication mechanisms are used to encode and share messages with selected agents, based on the learned topology. Each agent selectively receives and aggregates messages from its linked agents and updates its internal state accordingly. These updated hidden states are then used to instruct action, allowing agents to make informed decisions that reflect both their own observations and the shared knowledge received from others. Notably, the system maintains full decentralization - no global state or centralized controller is used at the execution time.
\vspace{-.2cm}
\section{Network Topology in MARL}
\label{sec:network}
In MARL, communication is not merely a matter of sending messages; it is fundamentally shaped by the topology of interactions among agents. The communication topology defines the directional flow of information among agents, establishing the temporal and structural parameters that govern inter-agent message exchange within the MAS. RIAL and DIAL \cite{foerster2016learning} pioneered the integration of communication within reinforcement learning, improving coordination through broadcasting. The fully connected network allows agents to share all their observations, but it quickly becomes impractical as agent populations grow. Hence, more researchers in MARL instead increasingly view communication as a learnable topology problem, where connectivity patterns are optimized alongside policies to balance information sharing efficiency with performance.
\vspace{-.45cm}
\subsection{Connection and Direction in the Topology}
Early approaches assumed full broadcasting, which guaranteed maximum observability but created redundant and noisy exchanges. To address communication inefficiency, IC3Net \cite{singh2018learning} introduced a binary gating mechanism that enabled agents to determine when to communicate based on contextual relevance, marking a shift from permanent to selective communication links. MAGIC \cite{niu2021multi} advanced this approach with a two-component architecture: a scheduler that determines both when and with whom to communicate, and a message processor that employs attention coefficients to weight incoming information during aggregation. This combination allows for both selective transmission and prioritized reception based on message relevance. HetNet~\cite{seraj2022learning} further extended these principles by incorporating structural heterogeneity, enabling specialized agent roles. These developments demonstrate how the communication link structure influences both system efficiency and functional organization in MAS coordination.
\vspace{-.5cm}
\subsection{Sequence of the Agents in the Topology}
While link direction defines the communication pattern, the order of information propagation determines how knowledge accumulates across agents over time. Real-world systems are rarely synchronous or symmetric. Therefore, the order in which agents update their policies significantly affects learning dynamics and final performance~\cite{wangorder}. In other words, sequential agent-by-agent optimization is sensitive to update order, as earlier-updated agents may dominate the coordination outcome. Building upon this insight, PMAT~\cite{hu2025pmat} explicitly optimized the order of action generation in MARL. By prioritizing influential agents to act earlier, PMAT allows critical information to propagate first through the network, leading to more effective coordination and improved task performance. Together, these works indicate that beyond the static structure of communication links, the temporal order of policy updates crucially shapes how knowledge accumulates among agents, directly influencing cooperation efficiency in complex MAS.
\vspace{-.5cm}
\subsection{Case Study}
The aforementioned work in MARL has shifted from debating the necessity of communication to optimizing its topology. This evolution necessitates a unified framework that integrates connectivity, directionality, and sequence. 

\subsubsection{DAG Topology}
To address this challenge, we develope a sequential coordination strategy that models communication topologies as Directed Acyclic Graphs (DAGs)~\cite{pearce2007dynamic}. In this design, agents are represented as nodes, and directed edges indicate the flow of information. The acyclicity property ensures that communication follows a causal order, preventing loops while enabling intent propagation: upstream agents share actions and observations that downstream agents can integrate into their represented observations and make decisions accordingly.
A key challenge in MARL is non-stationarity, where simultaneous policy updates prevent agents from adapting to each other's changing strategies in real-time, often destabilizing learning. The DAG-based topology overcomes this challenge by enabling sequential updating of the agent-by-agent policy optimization, which allows later agents to adapt to earlier ones when the execution order is appropriately structured. 
\begin{figure}[!t]
\centering
\includegraphics[width=.4\textwidth]{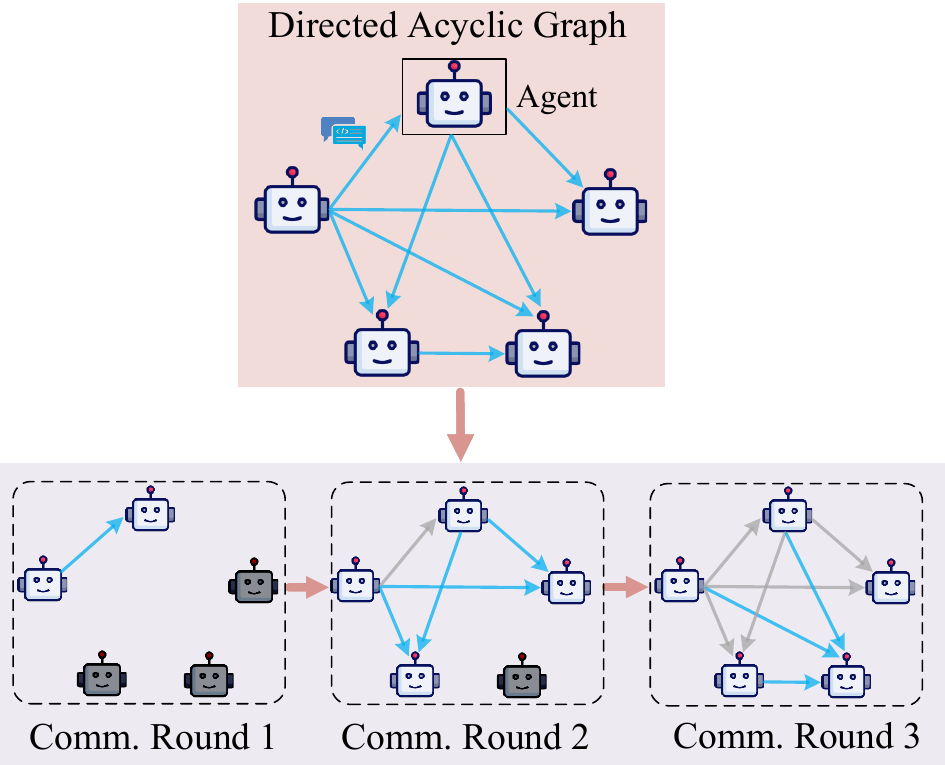}
\caption{Illustration of the DAGs topology order and depth. \textcolor{black}{The upper part shows the learned adjacency relationships between agents. The lower part demonstrates the resulting $3$ rounds of sequential communications among agents, derived from the DAG depth $d=3$ and sequential dependencies.}}
\label{fig:DAG}
\vspace{-.6cm}
\end{figure}
As illustrated in Fig.~\ref{fig:DAG}, the DAG topology is characterized by two fundamental properties:
\begin{itemize}
    \item \textbf{Order}: The order of a DAG represents the communication sequence among agents within the learned topology, deciding which agents act as information sources and which as receivers in the communication flow.
    \item \textbf{Depth}: The depth of a DAG, defined as the length of its longest path, determines the network's minimum communication rounds. It is the number of sequential steps required for a message to propagate from any agent to all reachable agents. Mathematically, the depth $d$ is given by $d=k-1$, where $k$ is the nilpotent index of its adjacency matrix $A$. The index $k$ is the smallest integer such that $A^k = \mathbf{O}$ (and $A^{k-1} \neq \mathbf{O}$).
\end{itemize}
Together, these properties ensure information flows systematically without central control~\cite{pearce2007dynamic} while characterizing the structural efficiency of information exchange in MAS.
\subsubsection{Performance Evaluation}
\textcolor{black}{To evaluate the proposed DAG-based approach, we examined two cooperative Grid World tasks: one with homogeneous agents (Predator-Prey, PP) and another with heterogeneous agents (Predator-Capture-Prey, PCP)~\cite{seraj2022learning}.} In both environments, $5$ agents operate in a $10 \times 10$ grid with a vision range of 1 and a maximum episode length of 80 steps. Each episode terminates early if agents successfully complete their task. In the PP environment, identical agents coordinate to track targets, while PCP requires specialized agents with distinct roles to cooperate in more complex capture sequences.
We compare our method against four baselines (MAGIC, MAT-dec, IC3Net, and Hetnet) using two key metrics: the number of steps required to complete the task and the communication overhead $C_{\text{comm}}$. For fair comparison, we define $C_{\text{comm}}$ as the average number of communications per episode. After training convergence, each algorithm is evaluated with the trained model across $100$ independent episodes. 
Each direct agent-to-agent transmission is counted as a single communication.
A broadcast to $n$ agents counts as $n$ communications. The total communication volumn is the aggregation of all rounds.

\begin{figure}[t]
  \centering
  \includegraphics[width=.45\textwidth]{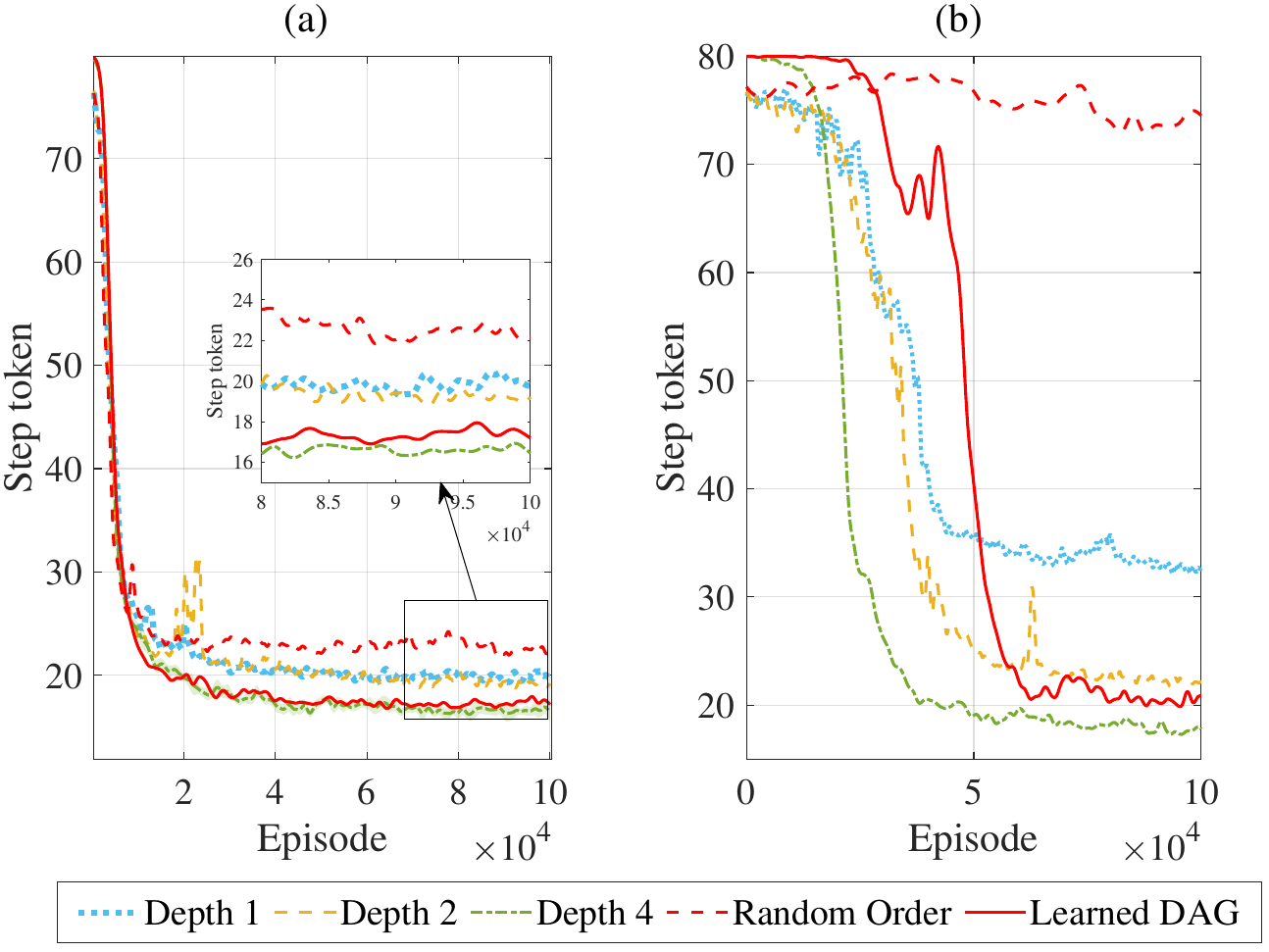}
  \caption{Validating the effectiveness of the proposed DAG topology: Impacts of depth and order on performance (a) PP (b) PCP.}
  \label{fig:3}
  \vspace{-.7cm}
\end{figure}

To validate the effectiveness of our learned DAG topology, a systematic study is performed to characterize its depth and order properties. The learned topology consistently has a depth of $2$ and is non-fully-connected.  
We then conduct ablation studies (Fig. \ref{fig:3}) to evaluate each property independently: (i) Depth Impact: Comparison against fully-connected DAG topologies with fixed depths (1, 2, 4). (ii) Order Impact: Comparison against DAG topologies that preserve the learned sparsity but use a shuffled communication order.
Our experimental results indicate several important findings. First, in the homogeneous PP environment, where communication primarily broadens agents' perceptual range, frequent information exchange helps reduce task completion steps. The depth of DAG topology shows moderate benefits, while variations in communication order produce only marginal differences-reflecting the functional equivalence of homogeneous agents.
In contrast, the heterogeneous PCP scenario demands tighter collaboration among specialized agents, where our DAG topology learning scheme demonstrates clear advantages. Deeper topology depth enables more rounds of information aggregation, substantially enhancing decision quality -- an effect particularly pronounced in this scenario with higher cooperation demands. Moreover, the learned order aligns closely with agents' functional roles, and disrupting it severely breaks information flow, leading to substantial degradation in task completion efficiency. Additionally, as shown in Table.~\ref{tab:network_comm_combined}, our learned topology enables efficient communication with less communication rounds and shorter average steps, indicating superior overall performance compared to all baselines.

Overall, this case study demonstrates that learning both the order and depth of communication is crucial for robust MARL, with their relative importance varying based on the degree of agent heterogeneity and task complexity. Table \ref{tab:network_comm_combined} summarizes these comparative results.

\begin{table*}[!t]
\centering
\caption{Performance comparison of network topology and communication information}
\label{tab:network_comm_combined}
\begin{tabular}{c c c c c !{\vrule width 1.5pt} l c c c c}
\toprule
\multicolumn{5}{c!{\vrule width 1.5pt}}{\textbf{Network Topology}} & \multicolumn{5}{c}{\textbf{Communication Information}} \\
\cmidrule(lr){1-5}\cmidrule(lr){6-10}
\multirow{2}{*}{\textbf{Grid World}} & \multicolumn{2}{c}{\textbf{Homo. PP}} & \multicolumn{2}{c!{\vrule width 1.5pt}}{\textbf{Heter. PCP}} 
& \multirow{2}{*}{\textbf{Traffic Junction}} & \multicolumn{2}{c}{\textbf{Original}} & \multicolumn{2}{c}{\textbf{Proposed Loss Adjust}} \\
\cmidrule(lr){2-3} \cmidrule(lr){4-5} \cmidrule(lr){7-8} \cmidrule(lr){9-10}
& Avg. step & $C_{\text{comm}}$ & Avg. step & $C_{\text{comm}}$ 
& & {Succ. Rate} & {Conv. Epoch} & {Succ. Rate} & {Conv. Epoch} \\
\midrule
IC3Net\cite{singh2018learning} & 25.62 & 343.31 & 52.63 & 842.08 
& CommNet\cite{sukhbaatar2016learning} & 0.715 & ~1500 & 0.998 (\textbf{+0.283}) & ~1200 (\textbf{-300}) \\
MAGIC\cite{niu2021multi} & 13.51 & 331.02 & 38.76 & 1257.42 
& G2ANet \cite{liu2020multi} & 0.961 & ~1300 & 0.986 (\textbf{+0.025}) & ~1050 (\textbf{-250}) \\
MAT-dec\cite{wen2022multi} & 30.12 & 601.74 & 73.61 & 1529.12 
& TarMAC\cite{das2019tarmac} & 0.794 & ~480 & 0.976 (\textbf{+0.182}) & ~1500 (+1020) \\
HetNet\cite{seraj2022learning} & 14.33 & 249.34 & 25.87 & 434.62 
& MAGIC\cite{niu2021multi} & 0.967 & ~750 & 1.000 (\textbf{+0.033}) & ~700 (\textbf{-50}) \\
\textbf{Proposed Topology} & 16.31 & \textbf{97.86} & \textbf{20.43} & \textbf{138.47} 
& IC3Net\cite{singh2018learning} & 0.877 & ~1500 & 1.000 (\textbf{+0.123}) & ~800 (\textbf{-700}) \\
\bottomrule
\end{tabular}
\vspace{-.5cm}
\end{table*}

\vspace{-.3cm}
\section{Communication Information in MARL}
\label{sec:comm}
The previous section studied the structure of communication -- its topology, timing, and propagation. We now turn to the content of communication, which is equally critical for efficiency in MARL~\cite{10342771}.
The design of communication messages determines whether exchanged signals are informative, redundant, or even harmful. In the literature, existing communication approaches can be systematically categorized into two predominant paradigms: (i) shared hidden states, where communication is realized by broadcasting internal representations, and (ii) attention-based mechanisms, where communication is filtered, targeted, and context-dependent.
\vspace{-.5cm}
\subsection{Hidden State Sharing}

Early work on communication for MARL emphasized direct sharing of hidden states. The principle is straightforward: an agent’s hidden state encodes both its observation history and policy-relevant information, making it a natural candidate for inter-agent communication.
DIAL\cite{foerster2016learning} introduced a framework where agents exchange real-valued messages derived from hidden states during training, enabling gradient flow across agents.
CommNet \cite{sukhbaatar2016learning} extended this idea by having agents broadcast their hidden states as continuous vectors, which can be aggregated through averaging to form communication inputs. This fully differentiable approach enabled communication to be trained alongside policies via backpropagation.
These approaches collectively highlight the benefits and limitations of hidden state sharing. While simple and effective in small cooperative tasks, broadcasting hidden states tends to produce redundant signals and communication bottlenecks as the number of agents grows. This motivates more structured mechanisms for communication message encoding.
\vspace{-.5cm}
\subsection{Attention Mechanism}
To overcome the inefficiency of indiscriminate hidden state broadcasting, attention-based mechanisms introduce selectivity into communication. These approaches can be understood from two complementary perspectives. First, targeted attention ensures that agents only interact with relevant peers rather than processing all incoming messages. For instance, TarMAC \cite{das2019tarmac} used a signature-query mechanism, where each agent attaches a learnable feature vector (signature) to its message that represents the message's content characteristics. Recipients generate query vectors based on their own states to compute attention weights with these signatures, enabling them to focus on messages that are most relevant to their current situation. This enables targeted communication channels that adapt dynamically to the task context.
Second, attention frameworks refine communication further by controlling the influence each message has on the recipient's decision-making process. G2ANet \cite{liu2020multi}, for example, employs a two-stage attention design: hard attention determines whether communication should occur or not(a binary decision about message relevance), while soft attention assigns continuous weights to different messages based on their estimated value to the current decision. These weights directly affect how strongly each message influences the recipient's internal state and subsequent actions, allowing agents to prioritize critical information while downplaying less relevant signals.

\vspace{-.5cm}
\subsection{Case Study}
Above discussion explains topology and message design; a logical extension of this analysis involves determining if agents utilize their communication channels to transmit concise, purposeful, and non-redundant information.

\subsubsection{Communication Information Efficiency}
To capture this dimension, we introduced two evaluation metrics into the training loop: 
\begin{itemize}
    \item \textbf{Information Entropy Efficiency Index  (IEI)}: The IEI quantifies how effectively agents encode task-relevant information in their communications. It is calculated as the average entropy across all communicating agents' messages. For each agent, we compute the entropy of its outgoing message by analyzing the distribution of values within the message vector. Lower IEI values indicate more concise and efficient information encoding, suggesting agents have learned to transmit only essential information without redundancy. This metric provides insight into the communication efficiency of the MAS relative to task performance. 
    \item\textbf{Specialization Efficiency Index (SEI)}: While IEI evaluates how compact the information is, the SEI measures how diverse the messages are across agents. SEI is calculated by first computing the pairwise cosine similarity between message vectors from different agents. Specifically, for any two agents $i$ and $j$, we determine the cosine similarity between their respective message vectors. These pairwise similarities are then averaged across all possible agent pairs to produce a single system-level similarity score. Lower SEI values indicate greater message diversity, suggesting that agents have developed specialized roles with complementary communication patterns. In contrast, higher SEI values reveal redundancy in communication, where agents transmit similar information regardless of their individual perspectives or functions. This metric helps quantify the degree of functional specialization that emerges during multi-agent cooperation, providing insight into whether agents contribute unique information to the collective task.
\end{itemize}

\subsubsection{Performance Evaluation}
\begin{figure}[!t]
  \centering
  \includegraphics[width=.45\textwidth]{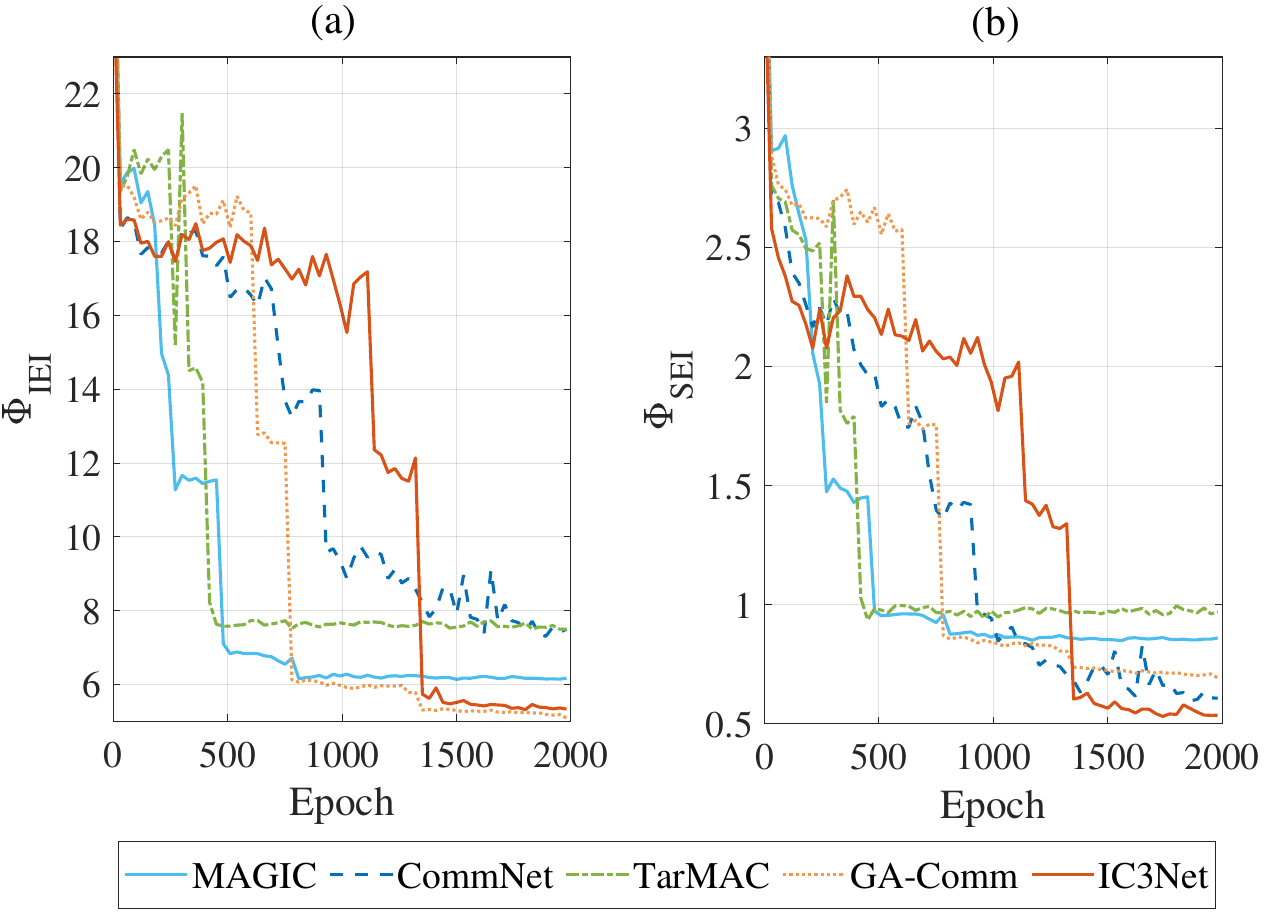}
  \caption{Comparison of $\Phi_{\text{IEI}}$ and $\Phi_{\text{SEI}}$ for different algorithms in the TJ environment.}
  \label{fig:TJ_diff_comm_round_Index2_Index3}
  \vspace{-.5cm}
\end{figure}
\textcolor{black}{
The Traffic Junction (TJ) environment~\cite{niu2021multi} consists of intersecting routes where cars (agents) with limited vision must communicate to avoid collisions. The cars enter from entry points with probability $p_\text{arrive}$ and follow randomly assigned routes. Each environment accommodates a maximum of $N_\text{max}$ cars, varying by difficulty level.
Cars occupy one cell per time step and can take "gas" or "brake" actions. Agent observations include previous action, route ID, and states of cells within vision range (set to 1). Collisions result in $-10$ reward per car, with an additional time penalty of $-0.01\tau$ per step (where $\tau$ is time since entry). Episodes are successful if no collisions occur.
In order to evaluate policy performance with and without the modifications to the training loss function, we validate communication efficiency at the easy level of TJ with two one-way roads on a $7\times7$ grid, two arrival points with two possible routes each and maximum of five agents ($N_\text{max}=5$, $p_\text{arrive}=0.3$). In our experimental setup, the training period consists of 2000 epochs, with each epoch containing 10 batches. Each batch processes 500 episodes, and each episode ends after 20 steps. The primary evaluation metric is the average success rate across episodes.}
Under the one-round communication environment, as shown in Fig.~\ref{fig:TJ_diff_comm_round_Index2_Index3}, all five algorithms -- CommNet, IC3Net, TarMAC, GA-Comm, and MAGIC -- displayed a common trend: both IEI and SEI started at relatively high levels and declined steadily over the course of training. This behavior suggests that agents initially relied on noisier and more redundant communication but gradually converged toward more compact and diverse message structures as their coordination strategies matured. 

Based on this observation, we directly integrate IEI and SEI into the training objective to explicitly encourage efficient and specialized communication. As summarized in Table~\ref{tab:network_comm_combined}, this modification leads to consistent improvements in both learning speed and/or final task performance in all algorithms evaluated. These results underscore the value of communication-aware objectives in MARL: rather than treating message exchange as a side effect, explicitly shaping communication through principled efficiency metrics can unlock stronger coordination in complex MAS environments.
\vspace{-.2cm}
\section{Challenges and Research Directions}
\label{sec:challenge}
Despite recent advances in designing communications in MARL, several challenges remain before these methods can be seamlessly deployed in practical MAS. In the following, we outline key obstacles and discuss research directions that may pave the way forward.
\vspace{-.4cm}
\subsection{Communication Environment Constrains}
A fundamental barrier arises from practical communication limitations such as restricted bandwidth, variable latency, and noisy channels. Current communication methods for MARL, including CommNet, IC3Net, and MAGIC, typically assume idealized communication with negligible transmission costs, which rarely holds in real deployments. Future research should therefore focus on adaptive communication strategies that can flexibly adjust communication frequency, message size, and routing topology according to network conditions. For example, integrating efficiency-oriented metrics such as the IEI into real-time communication scheduling could help agents prioritize task-critical signals while suppressing redundant transmissions under tight resource budgets.

\vspace{-.4cm}
\subsection{Interaction Environments}
Many benchmark tasks used today (e.g., Grid-World or Traffic Junctions) fail to capture the intricacies of real-world systems. Practical applications often involve cooperative, competitive, or mixed cooperative-competitive dynamics, each introducing unique communication requirements. In purely cooperative settings, the challenge lies in ensuring that information sharing scales with task complexity and observation horizon. In competitive environments, agents must balance communication for coordination within a team while simultaneously concealing sensitive information from adversaries. Mixed settings are even more demanding, as they combine elements of collaboration and opposition, requiring flexible protocols that adapt communication behaviors based on dynamic alliances or conflicts. Promising directions include hierarchical and role-based communication, where agents exchange task-relevant representations that generalize better across these diverse scenarios.
\vspace{-.4cm}
\subsection{Scalability with Agent Population}
In large-scale MAS, the central challenge shifts from communication efficiency to scalability. Dense all-to-all communication is impractical in large systems, and even attention-based methods face bottlenecks when populations grow to hundreds or thousands of agents. Future work should therefore investigate structured communication strategies, such as local broadcast, neighborhood-based message passing, or adaptive clustering. 
\vspace{-.4cm}
\subsection{Distributed Training and Execution}
Another major challenge lies in the gap between centralized training with decentralized execution and truly distributed execution. While CTDE has enabled progress, many implementations still rely on hidden centralization—such as a global critic or a central routing module -- which undermines full decentralization during execution. As systems scale, such centralized elements become impractical. Future research should therefore pursue genuinely distributed training paradigms, where agents learn with only partial information and local interactions. Promising directions include federated MARL, in which agents share gradients or model updates without revealing raw data, and asynchronous update schemes that mitigate synchronization bottlenecks. Embedding communication-efficiency measures into distributed optimization could also help stabilize learning and ensure that decentralized execution remains robust under real-world constraints.
\vspace{-.25cm}
\section{Conclusion}
\label{sec:conclusion}
MAS rely on effective coordination mechanisms for collective intelligence. We investigated two critical dimensions of communications in MARL: network topology and information efficiency. Our findings reveal that directed and sequential communication links significantly improve coordination quality while reducing required communication rounds, creating a more scalable approach for large systems. Additionally, our proposed metrics (IEI and SEI) quantify message compactness and diversity. Integrating them into training objectives accelerates policy convergence and improve task success. These results demonstrate that communications must be deliberately structured and optimized to adapt multi-agent learning dynamics. The fusion of adaptive topologies with information-efficient messaging may provide the foundation for robust MAS capable of addressing complex real-world challenges.

\bibliographystyle{IEEEtran}
\vspace{-.3cm}
\bibliography{IEEEabrv,mybib}
\end{document}